\documentclass[aps,prl,showpacs,reprint]{revtex4-1}
\usepackage{amsmath,amsthm,amssymb}
\usepackage{graphicx}
\usepackage{bm}
\usepackage{textcomp}
\usepackage{txfonts}

\begin{document}

\title{Controlling Sub-nm Gaps in Plasmonic Dimers using Graphene}

\author{Jan Mertens$^1$, Anna L. Eiden$^{2,1}$, Daniel O. Sigle$^1$, Antonio Lombardo$^2$, Zhipei Sun$^2$, Ravi S. Sundaram$^2$, Alan Colli$^3$, Christos Tserkezis$^4$, Javier Aizpurua$^4$, Silvia Milana$^2$, Andrea C. Ferrari$^2$, Jeremy J. Baumberg$^{1}$}
\email{jjb12@cam.ac.uk}
\affiliation{$^1$ NanoPhotonics Centre, Cavendish Laboratory, University of Cambridge, Cambridge, CB3 0HE, UK}
\affiliation{$^2$ Cambridge Graphene Centre, University of Cambridge, 9 JJ Thomson Avenue, Cambridge CB3 0FA, UK}
\affiliation{$^3$ Nokia Research Center, Broers Building, Cambridge CB3 0FA}
\affiliation{$^4$ CSIC-UPV/EHU and DIPC, Paseo Manuel de Lardizabal 5, 20018 Donostia-San Sebastian, Spain}


\date{\today}

\begin{abstract}
Graphene is used as the thinnest possible spacer between gold nanoparticles and a gold substrate. This creates a robust, repeatable, and stable sub-nanometre gap for massive plasmonic field enhancements. White light spectroscopy of single 80\,nm gold nanoparticles reveals plasmonic coupling between the particle and its image within the gold substrate. While for a single graphene layer, spectral doublets from coupled dimer modes are observed shifted into the near infra-red, these disappear for increasing numbers of layers. These doublets arise from plasmonic charge transfer, allowing the direct optical measurement of out-of-plane conductivity in such layered systems. Gating the graphene can thus directly produce plasmon tuning.
\end{abstract}

\maketitle
Precise separation of plasmonic nanostructures is of widespread interest, because it controls the build-up of intense and localised optical fields, as well as their spectral tuning.\cite{R1} Surface-enhanced Raman scattering relies on this plasmonic enhancement to enable identification of trace molecules captured within such gaps, down to femtomolar levels.\cite{R2,R3,R4,R5,R6,R7,R8} Plasmonic enhancement can also be used to improve the efficiency of photodetectors.\cite{R9} Similarly, nonlinear frequency conversion depends critically on the field enhancements, their spatial localisation, and their spectral resonances, all of which depend ultra-sensitively on the plasmonic gaps.\cite{R8} On the other hand fundamental processes like quantum tunnelling can be observed optically, but only for gap separations in the sub-nanometre regime.\cite{R10}
 
It is a major challenge to separate nanostructures precisely in the sub-nanometre regime because of poor structural control achievable on such length scales. Here we use a single layer of graphene (SLG) as a precise spacer between gold nanoparticles (AuNPs) and a Au surface to create stable sub-nanometre gaps down to 0.34\,nm. This allows us to optically probe the conductivity perpendicular to the graphene sheet, as well as to tune the plasmonic resonances by precise separation of NPs from the Au surface using flakes with increasing numbers of layers ($N$). Scattering from AuNPs placed directly on Au is compared with that of AuNPs separated from the Au substrate by flakes with increasing $N$. Splitting of the coupled plasmonic mode into a spectral doublet is observed due to charge transfer controlled by the SLG. Wide tuning of plasmonic resonances can thus become possible by mo\-du\-la\-ting the graphene dielectric response, for instance using gates. Such a robust nanoparticle-on-layered-material configuration forms a versatile platform for investigating a wide variety of plasmonic and monolayer functionalities and devices.

\begin{figure}[tb]
\includegraphics[width=\linewidth]{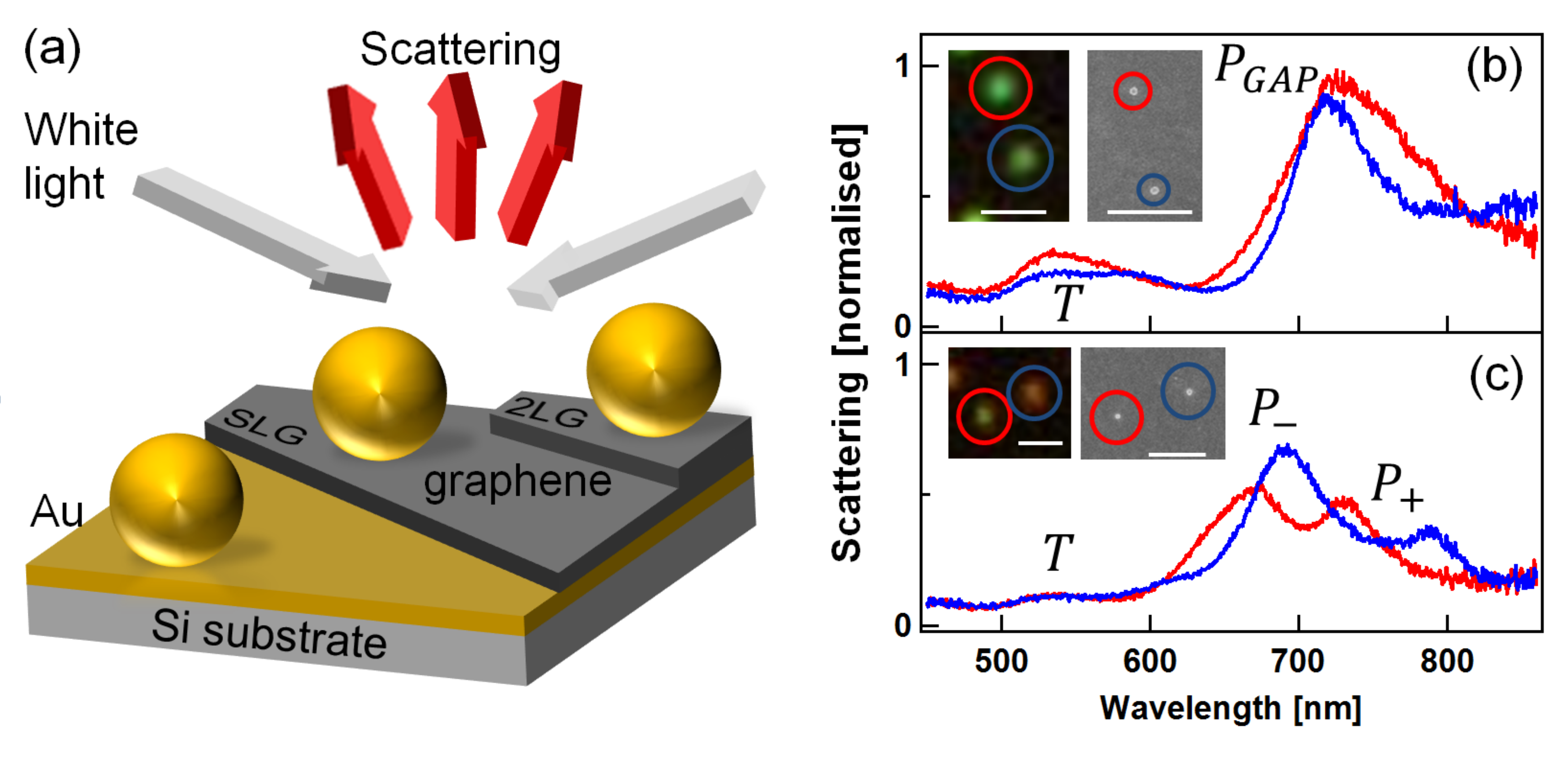}
\caption{
(a) Schematic of experiment: graphene placed on a thick Au substrate, and 80\,nm Au nanoparticles then deposited on top. Incident white light at high angles is scattered into the detector. (b,c) Dark-field single particle scattering spectra for (b) two AuNPs in direct contact with Au substrate, and (c) two AuNPs on CVD-grown monolayer graphene transferred onto Au. Insets show (left) dark-field images and (right) SEMs of same NPs. Scale bar is 1$\mu$m. }
\end{figure}
 
In our experiments a polished silicon substrate is coated with a 100\,nm Au film deposited by electron beam evaporation (Fig.1a). Graphene is prepared either via micromechanical cleavage of graphite or by chemical vapour deposition (CVD).\cite{R11,R12} The number of layers $N$ is determined by a combination of Raman spectroscopy\cite{R13,R14} and optical microscopy\cite{R15,R16}. Raman spectroscopy is also used to monitor the structural quality and doping of the graphene layers throughout the transfer process.\cite{R13,R14} Graphene is then transferred onto the Au film, layer by layer,\cite{R12} allowing us to deterministically control the spacer thickness. Transfer of CVD graphene is performed as follows:\cite{R11,R12} the initial SLG deposited on Cu is spin-coated with a thin layer of Poly(methyl methacrylate) (PMMA). Cu is then dissolved in a 3\%H$_2$O$_2$:35\%HCl (3:1 ratio) mixture, further diluted in an equal volume of deionized water. After all the Cu is dissolved, the remaining PMMA/graphene film is placed on the Au film, and PMMA is finally removed with acetone. Mechanically exfoliated graphene is transferred as follows:\cite{R11} a PMMA sacrificial layer is spin-coated on the flakes deposited onto si\-li\-con. The polymer-coated sample is then immersed in a NaOH solution, which partially etches the SiO2 surface releasing the polymer. Graphene sticks to the polymer, and can then be transferred. PMMA is eventually dissolved by acetone, thus releasing the flake.

Near-spherical 80\,nm AuNPs (BBI, citrate stabilised) are then self-assembled on a polished Si substrate treated with (3-aminopropyl)-triethoxysilane (APTES). The substrate is first dipped into a solution of 80\,nm AuNPs for 30\,s and then rinsed in deionised water to remove remaining unbound colloidal NPs. The self-assembled NPs are then transferred from the silicon substrate onto the graphene using the same PMMA transfer method previously described for the exfoliated graphene transfer. 
We characterise the scattering of white light from single AuNPs on the sample in areas that are covered with graphene flakes of increasing $N$. Single particles are identified using dark field microscopy. Unpolarised light from an incandescent source is focused on the sample using a 100x objective with a numerical aperture of 0.85, so that the sample is uniformly illuminated from large angles of incidence. The scattered light is detected with both a charge-coupled device (CCD) camera and also collected with a multimode optical fibre attached to a cooled spectrometer. The fibre has a core diameter of 50\,$\mu$m to give a collection spot with a diameter of $\simeq$\,500\,nm. Precise positioning of our AuNPs in the collection area is achieved using piezoelectric translation stages. The same AuNPs are also examined in a scanning electron microscope (SEM) to identify their shape. Direct correlation between SEM images and dark field spectra is necessary to correctly interpret the scattering spectra, since these depend strongly on the particle shape.

Dark field scattering spectra of four individual 80\,nm AuNPs are shown in Figs.1(b,c). The insets of these graphs show dark-field images (left) and SEM images (right) of the same AuNPs which are seen to be near-spherical. When the AuNPs directly contact the Au substrate (Fig.1b) two scattering modes, $T$ and $P_{GAP}$, are clearly visible for each particle, with resonance wavelengths of 530\,nm and 720\,nm respectively. The intensity of mode $P_{GAP}$ is 4 times stronger than T. In contrast when the AuNPs are separated from the Au film by a graphene monolayer three resonances are observed (Fig.1c). The resonance wavelength of the short wavelength mode T is unaffected by the SLG spacer, however in the near-infrared region, two modes $P_+$ and $P_-$ are detected. While the absolute spectral positions of the two modes for the two particles are not exactly the same, their spectral separation is comparable. Similar infrared spectral doublets are seen for all spherical AuNPs.

\begin{figure}[tb]
\includegraphics[width=\linewidth]{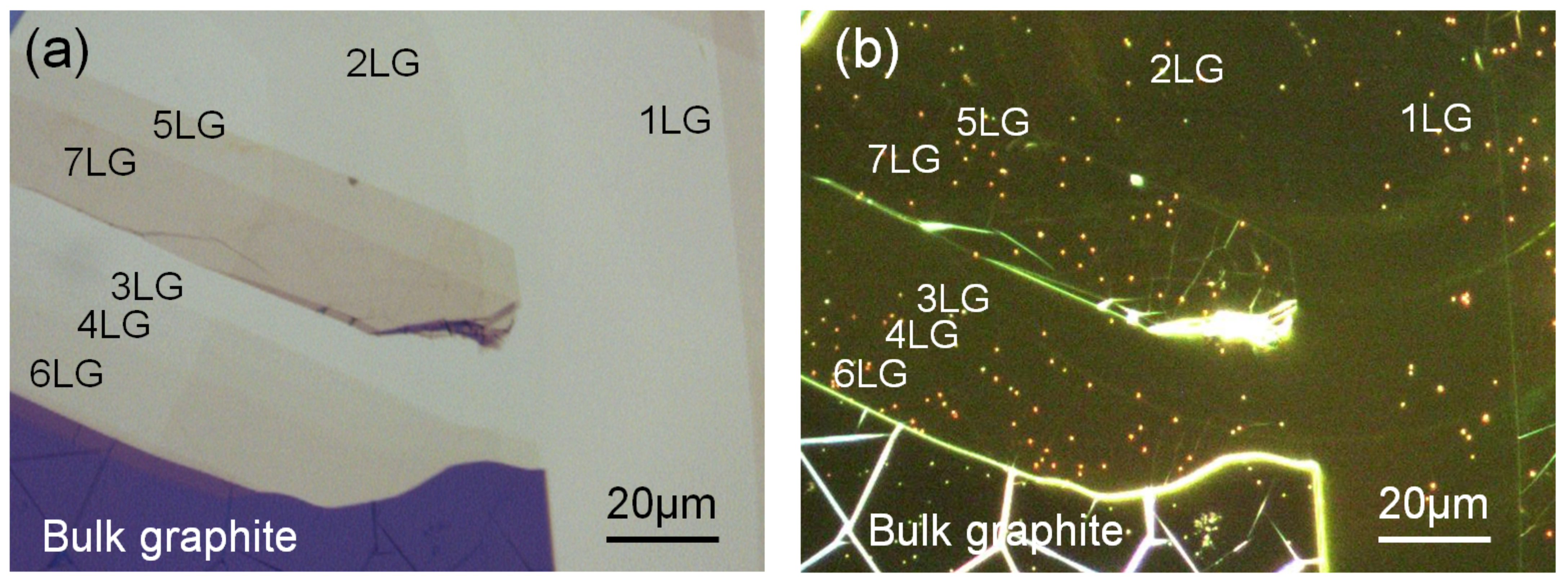}
\caption{
Exfoliated graphene on Au film. The number of graphene layers is indicated. (a) Bright field image of area before AuNP transfer. (b) Dark field scattering of the same area after AuNPs are transferred onto the graphene flakes. }
\end{figure}
 
As we discuss below, this doublet arises from the coupling of the normal dipolar plasmon across the AuNP and its image, with a gap plasmon highly localised in the SLG area directly underneath the NP. To prove this, we examine the dependence of the spectra on $N$. The AuNPs on top of the flakes are seen as bright spots in dark field images (Fig.2), and already show visibly the coupling effects since the AuNP spots on top of bulk graphite are spectrally shifted into the green compared to their red appearance in the 1-6LG regions. Fig.3(a) shows that the spectral doublet is only seen for the SLG spacer, while the infrared mode blue-shifts with increasing $N$ (Fig.3b). We emphasise that the spectral doublet when using a SLG spacer does not depend on how graphene is prepared (i.e. CVD or micromechanical exfoliation), and is reproducible across many AuNPs. While the spectra show some variability ($\pm$50\,nm) in the absolute spectral position of the doublets for small $N$, they become rather stable for $N>3$. This is expected from the extreme sensitivity of the plasmonic modes in the gap to the precise dielectric environment (for instance water).

\begin{figure}[tb]
\includegraphics[width=\linewidth]{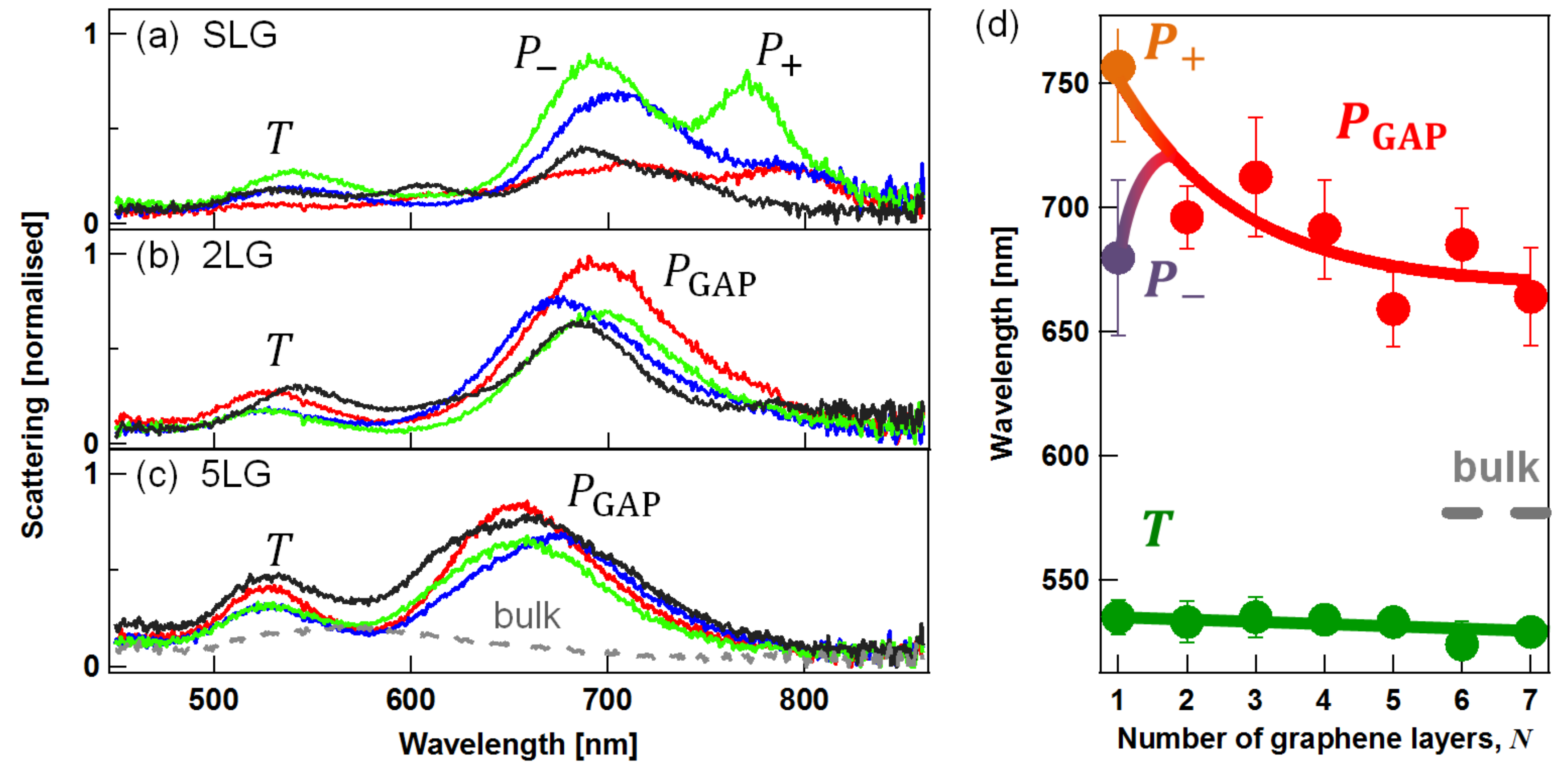}
\caption{
(a-c) Dark-field single particle scattering spectra of four AuNPs on SLG, 2LG and 5LG. Dashed grey curve (bottom graph) is scattering from AuNP on bulk graphite. (d) Resonance wavelength (averaged, variance given by the error bars) of the plasmonic $T$, $P_+$, $P_-$ and $P_{GAP}$ modes for increasing $N$. }
\end{figure}

To understand the origin of the doublet, we first show schematically the spectral evolution of the different plasmon modes associated with AuNPs on Au substrates spaced by conducting gaps (Fig.4a). This tuning map is based on experimental\cite{R17,R18} and theoretical\cite{R19,R20} studies of plasmons in NP dimers. While dimers are extremely hard to fabricate reliably, the NP-on-substrate geometry here produces an equivalent (but robust) system since solutions of the electromagnetic boundary conditions at the planar surface are equivalent to an image NP inside the metal. The NP sees its reflection in the mirror and couples to it, so that each charged region produces a dipole in conjunction with the oppositely-charged image NP. 

\begin{figure}[tb]
\includegraphics[width=\linewidth]{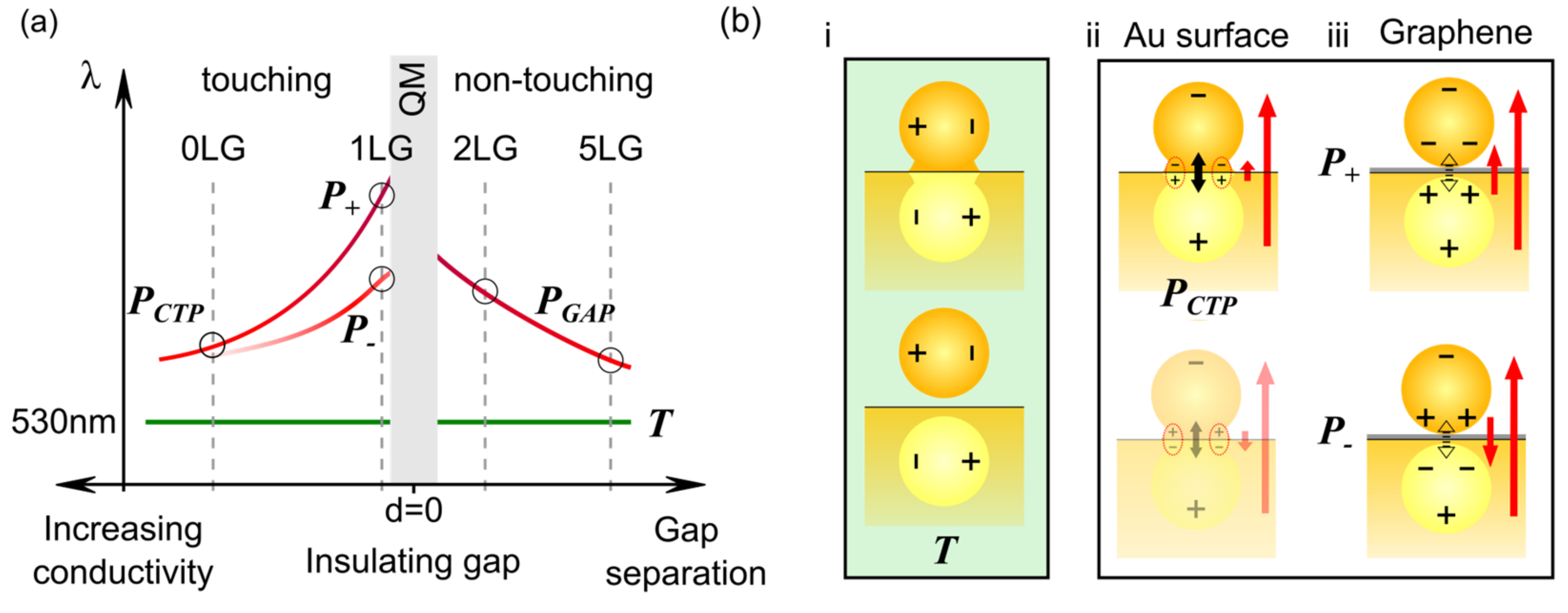}
\caption{
(a) Schematic spectral evolution of plasmon modes for AuNPs on Au with conducting gaps of different resistivity. Experimental data are labelled with the spacer thickness, indicated by the number of graphene layers (0-5LG). (b) Plasmonic charge distributions for AuNPs in contact with Au surface (i,ii) and when a SLG spacer is present (iii). Red arrows indicate induced dipole orientation and strength.}
\end{figure}

When the optical field is polarised in the plane of the substrate, it excites transverse plasmons across the AuNP ($T$) which are spaced far enough above the surface to produce only a weak coupling [Fig.4b,(i)] and a resonance close to that of individual AuNPs.\cite{R21,R22} Two other types of dipoles are driven by the component of the optical field along the dimer axis (excited by the high-angle incident light). One of these is the dipole resonance of the whole system, arising from the electric polarisation between the charge in the main body of the AuNP and its equal and opposite charge in the image NP. This global dipole plasmon (long arrows in Fig.4b) is present whenever the gap conducts and is the charge-transfer plasmon, $P_{CTP}$. The other dipole is located in the immediate vicinity of the gap (hence we term it the `gap plasmon'). This latter dipole (small arrows in Fig.4b, $P_{GAP}$) has a much smaller cross-section, since it is highly localised, and is extremely sensitive to the gap. These two longitudinal ($CTP$ and $GAP$) plasmons couple to each other to produce mixtures [Fig.4b(ii,iii)] in which they are aligned either parallel ($P_+=P_{CTP}+P_{GAP}$) or antiparallel ($P_-=P_{CTP}-P_{GAP}$), in a form of plasmon hybridisation closely analogous to ferromagnetic and antiferromagnetic dipolar coupling.\cite{R23} The strength of this strong coupling between $CTP$ and $GAP$ dipoles sets the splitting between the doublet states observed, and is controlled predominantly by the strength of the gap dipole.

The resonant wavelengths of the longitudinal plasmonic modes (Fig.4a, red lines) shift with gap separation and/or conductivity of the junction. Two main regimes are observed: the touching regime (left side), and the non-touching regime (right side), separated by a quantum mechanical tunnelling regime for Angstrom-scale gaps (labelled QM).\cite{R10,R19,R20} In the touching regime, the junction conductivity is high, thus charge can rapidly move between the AuNP underside and the Au film. This shorts out most of the dipole in the gap plasmon, leaving a single $CTP$ plasmon. In the non-touching regime, this charge transfer cannot occur, and the tuning and splitting of the two modes is set by the decreasing strength of the gap plasmon as the Au surfaces move further apart. The transverse modes (green lines) are not affected by the coupling, as expected from their charge distribution and consistent with our experimental observations (Fig.3b).
The red curves show that the energy splitting between $P_+$ and $P_-$ is largest when the gap is small (enhancing the gap plasmon cross section), and the junction is weakly conducting (avoiding discharging the gap plasmon). With no graphene in the gap, the Au atoms in the NP touching the Au film rearrange to form a significant neck [Fig.4b,(i)] with high conductivity. Such ductile necking between AuNPs is well known\cite{R24,R25,R26} and supported by our theoretical model (not shown) which indicates that the spectrum observed cannot arise for a perfect sphere resting on a plane. When a SLG spacer is used, the vertical resistance is sufficient to prevent discharge of the gap plasmon and both parallel-plasmon $P_+$ and antiparallel-plasmon $P_-$ modes are seen. Our theoretical models using literature values for the graphene dielectric constant\cite{R27,R28} show indeed that only a SLG spacer produces a doublet resonance. This confirms that the experimental gap width for the SLG spacer must be less than 0.4\,nm, (consistent with the 0.34\,nm SLG thickness). Increasing $N$ increases both the resistance (linearly with $N$), and also the gap width (which near-exponentially decreases the gap plasmon strength). Hence the dominant effect is to collapse the splitting between $P_+$ and $P_-$ modes, and blue-shift the gap plasmon. This is clearly seen in Fig.3(b), and marked as points 0-5LG on Fig.4(a).
 
The SLG thus acts in two ways. It forms a robust and perfectly controlled 0.34\,nm gap spacer between the Au layers with a high resistance compared to that needed to discharge the gap plasmon within half an optical cycle (1.5\,fs). However, it also prevents the wetting of AuNPs on the substrate, preserving their distinct near-spherical morphology on the non-wetting SLG.\cite{R29,R30} Au has a low adsorption on graphene so that the contact area of AuNPs on SLG is reduced compared to the direct contact of AuNPs on Au surfaces.\cite{R31,R30} A `digital' spacing step of 0.34\,nm is produced using flakes of increasing $N$,\cite{R15} allowing this geometry to be used as a test-bed for many plasmonic predictions as well as device functions.

The spectral separation of the doublet is sensitive to the gap width and vertical resistance through the SLG. Exact calculations of the charge transport from Au through SLG and back into Au require combinations of coupled Schrodinger-Poisson equations with electromagnetism.\cite{R10,R20} Vertical transport through a metal/SLG/metal structure can occur either via direct metal-to-metal tunnelling or via the graphene itself. Recent work\cite{R32,R33} supports the direct tunnelling model and reports vertical resistivities at low voltages of $\sim 5 \times 10^{-4} \, \Omega \, $cm$^2$ (at room temperature). For comparison, studies of tunnelling through metal/vacuum/metal systems using the Simmons formula\cite{R34} in the low-voltage approximation report tunnelling resistivities for a 0.3-0.4nm gap on the order of 10$^{-3} \, \Omega \, $cm$^2$ or lower.\cite{R35,R36,R37} On the other hand if transport occurs through graphene, a crude estimate of the vertical resistivity can be made from the inter-layer resistance in graphite, or from the average metal/graphene contact resistance in planar graphene devices. A realistic range would span from 10$^{-3}$ down to 10$^{-6} \, \Omega \, $cm$^2$  for certain metals.\cite{R38,R39} This range overlaps with estimates for the direct tunnelling above, and also with the value reported by Cobas et al.\cite{R32} The different spectral positions and separations of the doublets seen in Figures 1,3 result from this strong sensitivity to the transport in this local environment. The spectral response to light thus becomes a way to characterise the local vertical conductivity in graphene, and other two dimensional crystals obtained from the exfoliation of layered materials such as MoS2, WS2, BN, etc.,\cite{R12} as well as other semiconductors or more complex multilayer stacks. The enhanced local fields can already be seen in the surface-enhanced Raman scattering of the graphene in these gaps, which is enhanced by over two orders of magnitude for 633nm excitation, and over three orders of magnitude for 785\,nm excitation (see Supp.Info.). Given that the Raman laser spot size is 1\,$\mu$m in diameter, and the lateral width of the $P_+$ mode given by theory is 8\,nm (see Supp.Info.), this implies local Raman enhancements of greater than 10$^7$. Such large values offer opportunities for enhancing graphene-based photodetectors.\cite{R9}

In conclusion, we have shown that graphene acts as an ideal spacer for plasmonic nanostructures. In this NP-atomic layer-substrate configuration, graphene produces a stable and precise sub-nanometre separation of AuNPs from a gold surface and prevents coagulation of AuNPs with the gold substrate. A spectral doublet of the coupled plasmon resonance is observed in the near infra-red when NPs are separated from the Au substrate by a graphene monolayer. This doublet resonance corresponds to coupling between charge-transfer and gap plasmons mixed in parallel and antiparallel configurations. The spectral position of the resonances depends strongly on the resistivity of the junction between the NP and Au surface as well as the gap size, and strong shifts are observed with increasing numbers of layers. This work opens up the prospect for wide tuning of plasmonic resonances by modulating the graphene dielectric response, for instance using gates.


We gratefully acknowledge support from UK EPSRC grants EP/G060649/1, EP/G037221/1, EP/H007024/1, EP/G042357/1, EP/K01711X/1, EP/K017144/1, from ERC grants LINASS 320503 and NANOPOTS, from Nokia Research Centre, from a Royal Society Wolfson Research Merit Award,  EU grants RODIN, GENIUS, and CareRAMM. Jan Mertens acknowledges the support from the Winton Programme for the Physics of Sustainability.


\begin{thebibliography}{bib}
\bibitem{R23}  We note alternative terminologies also in use. For conducting gaps, the $P_+$ mode is termed the charge-transfer mode 1, and $P_-$ is termed the charge-transfer mode 2. In the insulating gap case, the gap plasmon mode is termed the bonding dipole plasmon (BDP).
\end{thebibliography}
\end{document}